\documentclass[12pt,preprint]{aastex}
\usepackage{graphicx}
\usepackage{natbib}
\newcommand {\ea} {{\it et~al.}}
\newcommand {\be} {\begin{equation}}
\newcommand {\ee} {\end{equation}}

\shorttitle{The 'blazar zone' in 3C454.3}
\shortauthors{Sikora \ea}

\begin{document}

\title{\object{3C454.3} reveals the structure and physics of its
'blazar zone'.}

\author{Marek~Sikora\altaffilmark{1},
Rafa{\l}~Moderski\altaffilmark{1}, 
Greg~M. Madejski\altaffilmark{2,3}} 

\altaffiltext{1}{Nicolaus Copernicus Astronomical Center, Bartycka 18, 00-716
Warsaw, Poland; \tt{sikora@camk.edu.pl}}
\altaffiltext{2}{Stanford Linear Accelerator Center, 2575 Sand Hill Road,
Menlo Park, CA 94025, USA}
\altaffiltext{3}{Kavli Institute for Particle Astrophysics and Cosmology,
Stanford University, Stanford, CA 94305}

\begin{abstract}
Recent multi-wavelength observations of \object{3C454.3}, in
particular during its giant outburst in 2005, put severe constraints
on the location of the 'blazar zone', its dissipative nature, and high
energy radiation mechanisms.  As the optical, X-ray, and millimeter
light-curves indicate, significant fraction of the jet energy must be
released in the vicinity of the millimeter-photosphere, i.e. at
distances where, due to the lateral expansion, the jet becomes
transparent at millimeter wavelengths.  We conclude that this region
is located at $\sim 10$ parsecs, the distance coinciding with the
location of the hot dust region.  This location is consistent with the
high amplitude variations observed on $\sim 10$ day time scale, 
provided the Lorentz factor of
a jet is $\Gamma_j \sim 20$. We argue that dissipation is driven by
reconfinement shock and demonstrate that X-rays and $\gamma$-rays are
likely to be produced via inverse Compton scattering of near/mid IR
photons emitted by the hot dust.  We also infer that the largest
gamma-to-synchrotron luminosity ratio ever recorded in this object --
having taken place during its lowest luminosity states -- can be
simply due to weaker magnetic fields carried by a less powerful jet.
\end{abstract}

\section{Introduction}
Multi-wavelength coverage of recent activity of quasar
\object{3C454.3} provided exceptional data to investigate the
structure and physics of its blazar zone. Prior to year 2000, this
object spent most of its time in the low, relatively quiescent state.
Starting in 2000, \object{3C454.3} entered a highly active state,
changing optical flux by a factor tens on time scales of a few months
\citep{Fuhrmann06,Villata06}.  The most powerful event took place in
the middle of 2005.  This event was monitored also in the X-ray bands
(Swift/XRT/BAT: \citealt{Giommi06}; INTEGRAL: \citealt{Pian06};
Chandra: \citealt{Villata06}), and at millimeter wavelengths
\citep{Krichbaum07}.

These data allow a construction of quasi-simultaneous broadband
spectrum around the outburst peak.  As is the case for other blazars,
the spectrum is composed of two humps, the lower energy one produced
via synchrotron mechanism and peaking in the far-infrared band, and
the higher energy one most likely generated by inverse-Compton process
and peaking in the $\gamma$-ray band.  The lack of coverage of the
event by $\gamma$-ray observatories does not allow us to determine the
luminosity of the high energy component.  Nevertheless, X-ray data
suggest that luminosity ratio of the high- to the low-energy
components was much smaller during the outburst than during low states
monitored in $\gamma$-rays by CGRO
\citep{Mukherjee97,Hartman99,Zhang05}.

This difference was theoretically investigated by \citet{Pian06} and
by \citet{Katarzynski07}.  \citet{Pian06} suggested that during the
low states the blazar zone is located inside the broad line region
(BLR) and that high energy spectra are produced by the External
Radiation Compton (ERC) process involving scattering of broad line
photons \citep[via scenario described in][]{Sikora94}, while during
the 2005 outburst the dissipation zone moved outside the BLR where the
ERC becomes inefficient.  In such a model, production of the optical
outburst doesn't require increase of a jet power.  Similarly, in the
scenario proposed by \citet{Katarzynski07} the jet power is constant,
but the drop of luminosity of the high energy component is explained
by decrease of the Lorentz factor.

The idea of the constant jet power might be challenged by the most
recent optical outburst which in July 2007 was also detected in
$\gamma$-rays by AGILE \citep{Vercellone07}.  Bolometric luminosity of
this outburst was 4-5 times larger than bolometric luminosity during
the low optical states and the radiative output was strongly dominated
by the $\gamma$-ray flux.  The currently available millimeter-band light
curves \citep{Krichbaum07} do not indicate any significant delay of
the millimeter flux after the bolometric flux as inferred from the
infrared and optical data presented in \citet{Bach07}.  All of
the above motivated us to investigate a different scenario, with the
origin of the high energy peak involving ERC with IR seed photons and
operating in the vicinity of the millimeter photosphere of the source.
Basic assumptions of the scenario are described in \S\ref{sec:assum};
results of modeling of the broadband spectrum of the 2005 outburst are
presented in \S\ref{sec:outburst}; explanation of the large
$\gamma$-ray dominance in the low optical states is provided in
\S\ref{sec:model}; and the main results are summarized in
\S\ref{sec:discussion}.

\section{Model assumptions}
\label{sec:assum}
\subsection{Location of the blazar zone\label{sec:location}}
Optical and millimeter light-curves show that the '2005 outburst' of
\object{3C454.3} was actually preceded by a long term gradual increase
in flux which started in August 2004 and continued until the middle of
2005 \citep{Villata06,Krichbaum07}.  The optical flux reached maximum
around May 9, then dropped very rapidly, but this drop was associated
with several local ``wiggles''.  The millimeter light-curve reached
maximum about 18 days later and continued at that level for $\sim 3$
months with fluctuations on a time scale of $\sim 10$-days.  The
outburst ceased by the August/September 2005.  The lack of a high
luminosity plateau in the optical light curve suggests that the
millimeter outburst lags the optical one by $\sim 3$ months.  However
no such long delay is seen in the growing part of the outburst.
Furthermore, the optical spectrum is steep and very variable which
makes the optical flux a very poor tracer of the {\sl bolometric}
luminosity.  The latter, according to data presented by
\citet{Bach07}, presumably reached the maximum (with the peak located
in the far IR) by the end of June 2005, roughly in the middle of the
millimeter plateau.  This, coupled with large millimeter luminosities
which require {\sl in situ} energy dissipation rate that is comparable
to the rate required to account for optical emission -- and similar
short term variability time scales in both spectral bands -- suggest
that regions of the optical and millimeter emission are not spatially
detached.

If the above is indeed the case, it is possible to make unambiguous
estimates of the location of the blazar zone (with respect to the
central black hole) based on the variability time scales, and this in
turn can be verified by using millimeter data and calculations of the
synchrotron-self-absorption opacity of the source.  Since the spectral
slope measured in the millimeter band during the outburst is typically
within the range $0.0 < \alpha_{mm} < 0.5$, the blazar zone is
expected to be partially opaque at these wavelengths.  The resulting
size of the source $R_{mm}$ and its distance from the center $r_{mm}$
depend on the specific model parameters and for those presented in
Table~\ref{tab:param} are calculated to be $R_{mm} \sim 0.5$ pc and
$r_{mm} \sim 9$ pc (see Appendix~\ref{app:photo}).

\subsection{Dissipation scenario}
While it is relatively well-established that the endpoints of most
quasar jets correspond to ``hot spots'' presumably involving terminal
shocks, there is no consensus regarding the mechanism responsible for
the energy dissipation within the flow and in particular in the blazar
zone. Most popular, presumably because it is the easiest to treat
quantitatively, is the internal shock scenario.  In accordance with
this scenario, jets are radially inhomogeneous both in density and
velocity and shocks are formed due to collisions between jet portions
propagating with different Lorentz factors \citep{Sikora94,Spada01}.
Internal shock scenario is attractive for blazars because predicts
parallel polarization (electric vector position angle, EVPA, parallel
to the jet) of the synchrotron radiation, in agreement with
observations in the optical, infrared, and millimeter bands
\citep{Impey91,Stevens96,Nartallo98,Jorstad07}.  This prediction is
independent of whether magnetic field is dominated by the toroidal
component determined by poloidal electrical currents or by turbulent
magnetic fields compressed in the transverse shocks \citep{Laing81}.
However, internal shocks are known to dissipate energy very
inefficiently: modulation of a jet Lorentz factor by at least a factor
of $4$ is required to reach a few percent of efficiency.

More promising dissipative scenario involves reconfinement shocks
\citep{Komissarov97,Sokolov04}.  Such shocks keep pressure balance
between the jet and its environment and are formed everywhere where
density gradient of the external medium departs from the longitudinal
density gradient in a jet.  On sub-parsec scales the environment is
too weak to affect dynamically powerful jets, but at parsec and larger
distance, the interaction of the jet with its environment is
sufficiently strong to modify the opening angle and, in the case of
non-axisymmetric external matter density distribution, also the
direction of propagation \citep[see e.g.][]{Appl96}.  Reconfinement
shock scenario provides interesting constraints on the structure and
intensity of magnetic fields.  In such shocks compression of chaotic
magnetic fields leads to the perpendicular EVPA, but if magnetic field
intensity is dominated by the toroidal component, the EVPA is parallel
to the jet, in agreement with observations.

\subsection{Radiative mechanisms and model input parameters}
Basic radiative processes in relativistic jets are known to be the
synchrotron mechanism and the inverse-Compton process.  The latter
involves scatterings of both 'internal' synchrotron photons (the SSC
process) and 'external' photons (the ERC process).  The ERC is
expected to dominate strongly over the SSC provided radiative
environment is strong and jets are highly relativistic
\citep{Dermer92,Dermer93,Sikora94,Blandford95}.  At parsec distances,
corresponding to the likely location of the blazar zone in
\object{3C454.3} (see \S\ref{sec:location}), the external diffuse
radiation field is dominated by near/mid infrared radiation of hot
dust \citep[and refs. therein]{Cleary07} and therefore such dust is
very likely to provide the dominant source of seed photons for the
inverse-Compton process \citep{Blazejowski00,Arbeiter02}.  This is in
fact the scenario suggested for the origin of the high-energy peak in
MeV blazars \citep[a class of blazars also encompassing 3C454.3;
  see][]{Sikora02} and is the scenario adopted below.

To reproduce the broadband spectrum of radiation produced in the blazar
zone, we apply the numerical model BLAZAR \citep{Moderski03}, updated
for the treatment of the Klein-Nishina regime \citep{Moderski05}.
Originally, the model was designed to compute radiation spectra
assuming the internal shock scenario, but noting that steady-state
radiation can be superposed from a sequence of moving sources which
all radiate within the same distance range, the model can be used also
to approximate radiation production by the standing reconfinement
shock.

Possibly the most significant simplification of our model is that we
do not consider real geometry and kinematics of the reconfinement
shock, adopting instead the uniform injection/acceleration of
relativistic particles within the conically diverging zone.  The
details of the physics of reconfinement shocks and in particular of
particle acceleration are still not known, and it is even unclear
whether the dissipation process and particle acceleration involve just
the reconfinement shock or some sort of a hybrid model incorporating
internal shocks amplified in the reconfinement zone
\citep{Komissarov97,Sokolov04}.

The following input parameters are used in our model:
\begin{itemize}
\item radial extension of the blazar zone, $\Delta r$, and the
  distance of its inner edge from the center, $r_0$;

\item the jet Lorentz factor, $\Gamma_j$, and its opening (half) angle
  $\theta_j$;

\item magnetic field intensity, $B=B_0 \times (r_0/r)$;

\item the electron injection function, $Q = K \gamma^{-p}$ for
  $\gamma_{min} < \gamma < \gamma_{max}$;

\item energy density of the diffuse component of hot dust radiation,
  $u_{IR} = u_{IR,in} \times [1 + (r/r_{in})^2]^{-1}$, where $r_{in}$ is
  the inner edge of the hot dust region, $u_{IR,in} \sim \xi_{IR}
  L_{disk} / (4 \pi r_{in}^2 c)$, $L_{disk}$ is the accretion disk
  luminosity, and $\xi_{IR}$ is the fraction of the disk radiation
  reprocessed by dust into infrared radiation;

\item energy of the seed photons at thermal peak in $\nu L_{\nu}$
  vs. $\nu$ diagram, $h\nu_{IR} \simeq 3.92 \, k T$, where $T=
  T_{in}(r_{in}/r)^{1/2}$, and $T_{in} = (L_{disk} /(4\pi \sigma_{SB}
  r_{in}^2 c))^{1/4}$.

\end{itemize} 

Values of these parameters are determined by our model assumptions and
by relations between these parameters and observables.  The latter, in
the form of approximate formulas, are presented in
Appendix~\ref{app:approx}.  Analytically estimated parameters are used
to start an iterative procedure to fit numerically the observed
spectrum.  Because the 2005 outburst was not observed in the
$\gamma$-ray band and because of uncertainties regarding distribution
and opacity of the hot dust, the set of input parameters cannot be
determined uniquely.  This in particular concerns the value of the jet
Lorentz factor.  We assumed $\Gamma_j = 20$.  Such a large value
allows us to avoid softening of the X-ray spectrum by contribution of
the SSC process in the soft/mid X-ray bands.  Such a large value of
$\Gamma_j$ is also implied when we adopt the assumption of domination
of the toroidal magnetic component over the turbulent one.  The
$\Gamma_j = 20$ is larger than that deduced from the VLBI observations
of the superluminal expansion \citep[see][and
  refs. therein]{Jorstad01}, but the latter can be underestimated due
to not taking into account effects of the divergence of a jet
\citep{Gopal06}.

\section{Modeling the 2005 outburst\label{sec:outburst}}
\begin{figure}
\centering
\includegraphics[width=0.95\textwidth,bb=50 50 555 495]{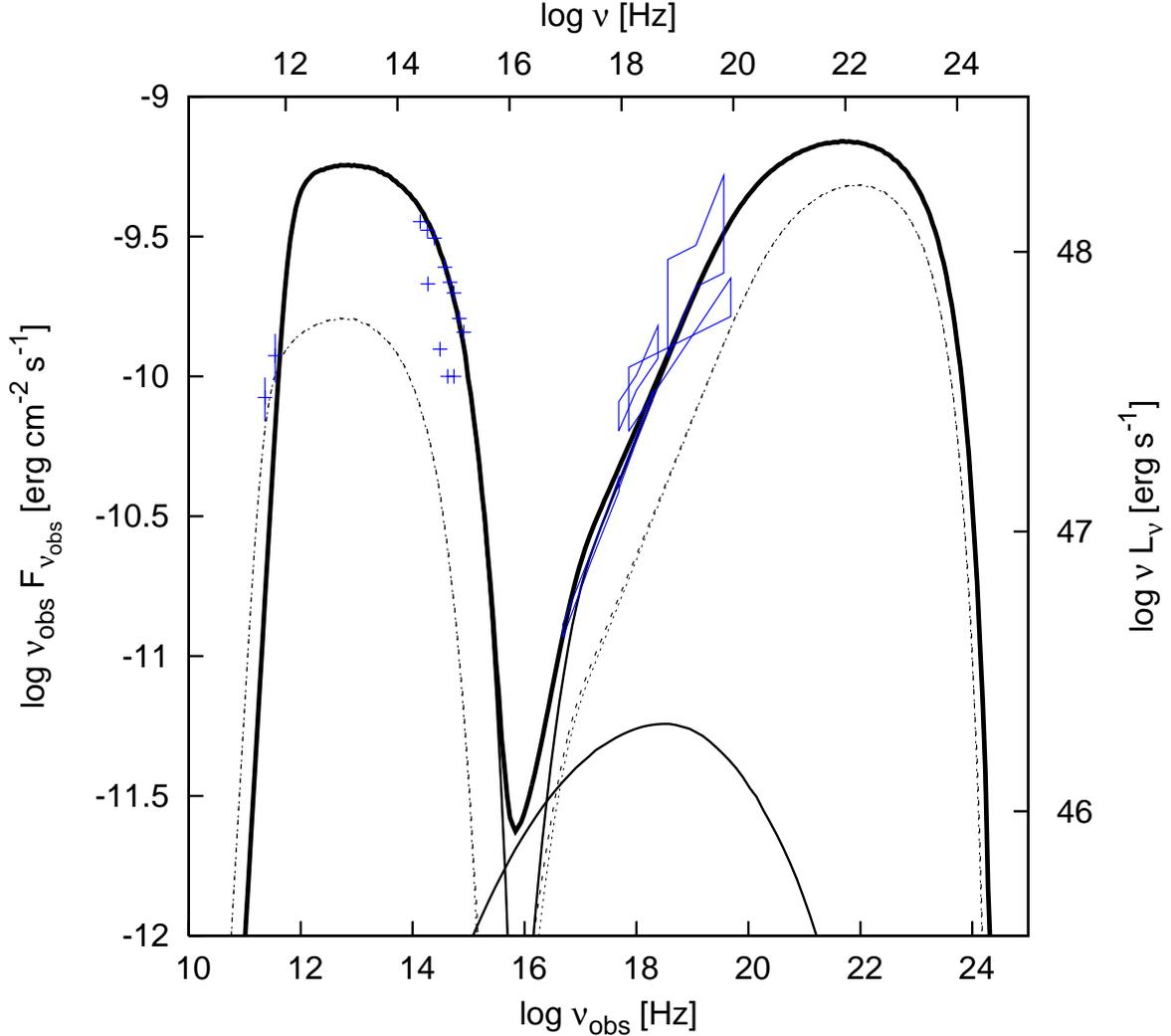}
\caption{Data points show the broadband spectrum of 3C454.3 at the
  epoch of the optical peak during the 2005 outburst. Infrared data
  points at $1\,$mm and $3\,$mm and upper optical data points are from
  IRAM telescope and WEBT campaign, respectively, and were reported
  together with Chandra data in \citet{Villata06}.  Lower optical data
  points from REM telescope and Swift data are taken from
  \citet{Giommi06}.  Integral data are from \citet{Pian06}. Continuous
  lines show our preferred models obtained using the $BLAZAR$ code
  \citep{Moderski03}.  Thick, solid line shows the model accounting
  for the broad-band data during the optical peak of the outburst
  (Model 1 in Table~\ref{tab:param}); the thin, solid lines indicate
  various components of the spectrum and illustrate that the SSC
  component is relatively weak.  Dashed lines show the model spectrum
  produced at a distance twice as large as the thick solid line, and
  are intended to illustrate the emission at the millimeter
  photosphere (Model 2 in Table~\ref{tab:param}).  Model parameters
  are given in Table~\ref{tab:param}.}
\label{fig:model1}
\end{figure}

Results of modeling of the spectrum observed in May 2005, when the
optical flux was at its maximum are shown in Fig.~\ref{fig:model1} and
input and output parameters are specified in Table~\ref{tab:param}.
As it is apparent, the entire spectrum can be reproduced using a
single-power-law for electron injection function, with a slope index
$p=2$.  X-ray spectrum is produced by electrons which cool on a time
scale longer than the blazar-zone crossing time and therefore this
results in the slope $\alpha_X = (p-1)/2 \simeq 0.5$.  Synchrotron
spectrum is produced in the fast cooling regime and results in the
slope $\alpha_{syn} = p/2 \simeq 1.0$, but in the optical band it
significantly steepens due to high energy cutoff in the injection
function. It hardens at the millimeter wavelengths due to synchrotron
self-absorption.

Our results show that even a very moderate energy density of the dust
radiation is sufficient to provide strong domination of the ERC
luminosities over the SSC luminosities.  This is due to a large value
of $\Gamma_j$ and strong dependence of the $L_{ERC}/L_{SSC}$ ratio on
$\Gamma_j$. The spectrum shown in Fig.~\ref{fig:model1} is obtained
for an active zone enclosed within a distance range $10^{19} - 2
\times 10^{19}\,$cm.  Jet within this distance range is opaque at
millimeter wavelengths.

In order to get spectrum with the observed slopes and fluxes in the
millimeter band, it is necessary to assume a larger distance of the
blazar zone and smaller optical luminosities.  In
Fig.~\ref{fig:model1} we show the broadband spectrum produced within a
distance range $2 \times 10^{19} - 4 \times 10^{19}\,$cm.  Optical
luminosity is smaller there by a factor $\sim 5$, but assuming that
magnetic energy flux is proportional to the flux associated with
matter flow, it was possible to accommodate this by decreasing the
electron injection function by only a factor of $2$ (see parameters in
Table~\ref{tab:param}).  Optical luminosity produced within this
distance range corresponds with optical fluxes recorded during the
millimeter-plateau period.  Results from Fig.~\ref{fig:model1}
indicate that most powerful portions of the jet start to dissipate
energy closer to the center than the less powerful ones, but energy
dissipation extends, albeit with a decreasing efficiency, up to the
region where the plasma becomes transparent at millimeter wavelengths.

\section{Modeling different spectral states\label{sec:model}}
\begin{figure}
\centering
\includegraphics[width=0.95\textwidth,bb=50 50 555 495]{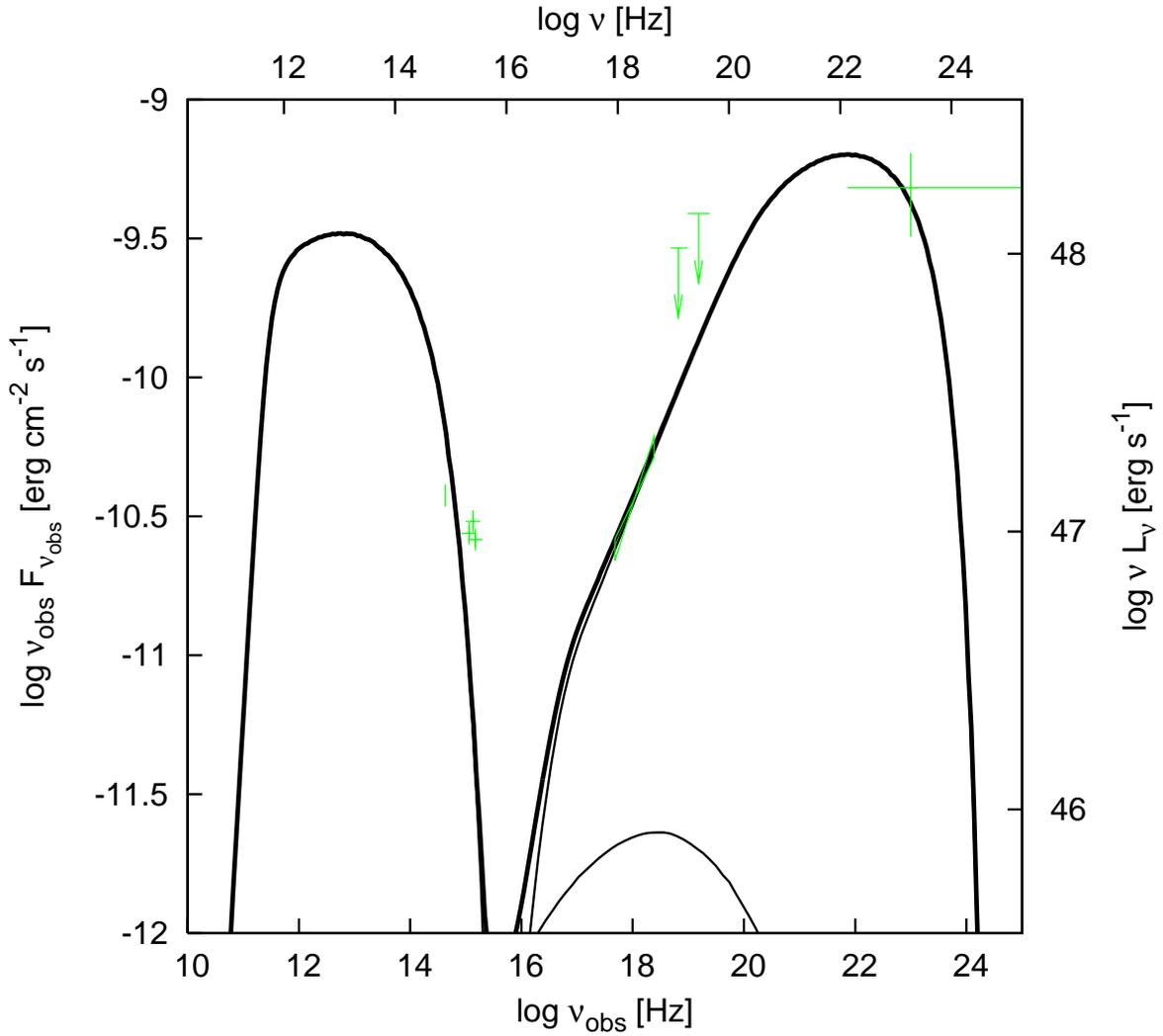}
\caption{Broadband spectral observations of 3C454.3 during the 2007
  outburst.  Tuorla Observatory optical data point and Swift UV and
  X-ray data are taken from \citet{Ghisellini07}.  Agile point comes
  from \citet{Vercellone07}.  Model illustrated as a solid line has
  parameters given in Table~\ref{tab:param} as Model~3.}
\label{fig:model3}
\end{figure}

\begin{figure}
\centering
\includegraphics[width=0.95\textwidth,bb=50 50 555 495]{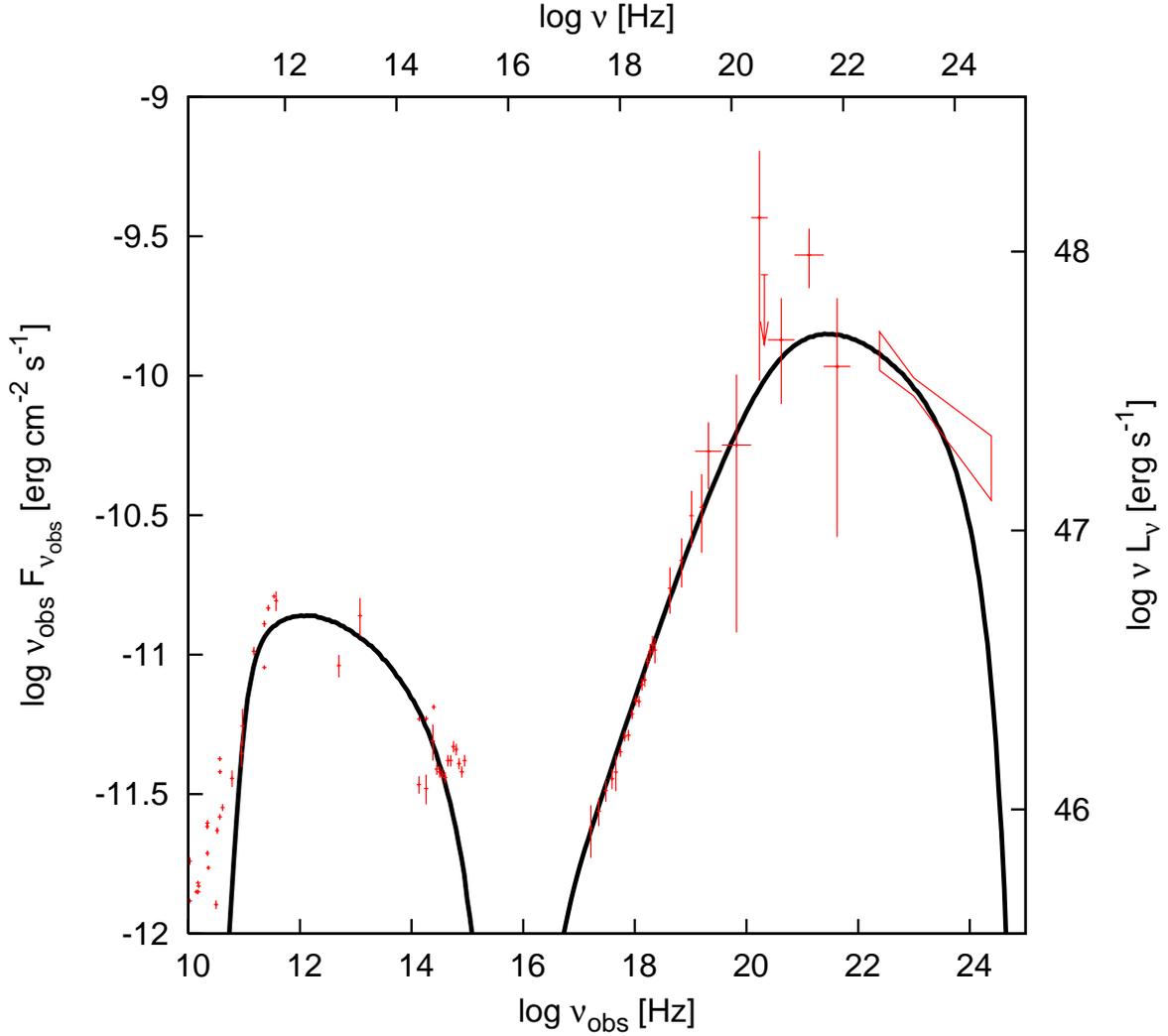}
\caption{Broadband spectrum during the low state in the 'CGRO epoch'.
  All data points below $10^{16}\,$Hz are from NASA Extragalactic
  Database.  BeppoSAX data come from \citet{Tavecchio02}, while CGRO
  OSSE, Comptel and EGRET data are from \citet{McNaron95},
  \citet{Zhang05}, and \citet{Hartman99}, respectively.  Model
  accounting for those data, with parameters given in
  Table~\ref{tab:param} (Model~4) is plotted as a solid line.}
\label{fig:model4}
\end{figure}

Important observable characterizing the double-hump spectra of blazars
is the luminosity ratio of the high energy component to the low energy
component.  If production of a high energy component is dominated by
the ERC process, then this ratio is $L_{ERC}/ L_{syn} \sim \Gamma_j^2
u_{IR}/u_B'$, where $u_B'$ is energy density of the magnetic field in
the blazar zone of a jet.  Noting that energy flux of magnetic field
in a jet is $L_B \simeq c u_B' \pi R^2 \Gamma_j^2$ and $u_{IR} =
\xi_{IR} L_{disk} /(4\pi r^2 c)$, and assuming that $L_B \propto
L_{jet}$ and $\theta_j = R/r \sim 1/\Gamma_j$, this ratio is
\be {L_{ERC} \over L_{syn}} \propto {\Gamma^2 \xi_{IR} L_{disk} \over
L_{jet}} \label{eq:lum}\ee
Hence, for a fixed disk luminosity, luminosity ratio of the two components
depends mainly on three parameters, $\Gamma_j$, $\xi_{IR}$, and
$L_{jet}$.  All of them can be a function of a distance in a jet, and
$\Gamma_j$ and $L_{jet}$ can additionally vary with time.  With our
basic assumption that the blazar zone is related to the location of
the reconfinement shock and that this location is not changing
significantly with time, changes of the luminosity ratio from the
epoch to the epoch can be just a function of $L_{jet}$ and $\Gamma_j$.
We demonstrate in Fig.~\ref{fig:model3} and \ref{fig:model4} that
spectra of \object{3C454.3} taken at two epochs, during the outburst
and during the quiescent phase, can be reproduced just by assuming
changes in $L_{jet}$ and some modifications in the shape of the
injection function. From inspection of these spectra (including
Fig.~\ref{fig:model1}), it is apparent that differences between
synchrotron luminosities at different states are much larger than
differences between bolometric luminosities.  This results from the
fact that for $L_{ERC} > L_{syn}$, $L_{ERC} \sim L_{bol} \propto
L_{jet}$, and when this is combined with the Eq.~(\ref{eq:lum}), 
it gives $L_{syn} \propto L_{jet}^2$.
 
\section{Discussion and conclusions\label{sec:discussion}}
We demonstrated in this paper that broadband spectra of
\object{3C454.3} can be reconstructed assuming that they are produced
at distances $r \sim 3-9$ parsecs.  By the end of this distance range
the jet becomes transparent at millimeter wavelengths.  Blazar
activity historically has been defined via observations in the
IR/optical bands, while ``blazar-zone'' is often considered to be
located deeply within the millimeter photosphere.  However, the
optical and millimeter light-curves seem to indicate a significant
overlap of the blazar-zone with a region where the jet becomes
transparent at millimeter wavelengths (see \S\ref{sec:location}).
This is further supported by very large millimeter luminosities which
require high, {\it in situ}, dissipation rate of energy, and is
consistent with time scales of the fastest high amplitude variations,
of the order of $10$ days in both spectral bands. Furthermore, at such
distances the co-spatial model self-consistently incorporates
production of X- and $\gamma$-rays, via scatterings of near/mid IR
photons emitted by hot dust.

It should be emphasized here that the input-parameter set for ERC
models is not unique and that high energy spectra can be reproduced
also by scattering of broad emission photons if taking place in the
sub-parsec region.  However, then the high energy non-thermal
radiation should be accompanied by bulk-Compton features
\citep{Sikora00,Moderski04,Celotti07}, which so far have not been
observationally confirmed.  Their lack or weakness can be explained by
assuming that in the sub-parsec region jet is still in acceleration
phase and the blazar zone is located at larger distances
\citep{Kataoka07}.

We identify the ``blazar zone'' with a reconfinement shock.  That,
together with optical polarization data imply domination of the
toroidal magnetic field over chaotic/turbulent magnetic fields.
However it should be noted that domination of the toroidal component
doesn't necessary indicate the domination of the Poynting flux over
the matter energy flux. It is very likely that the conversion of the
Poynting flux dominated jet into matter dominated jet -- and hence the
jet acceleration process -- are accomplished on sub-parsec scales
\citep{Sikora05,Komissarov07}.  Similar conclusions are reached by
\citet{Jorstad07}, following multi-waveband polarimetric observations
of 15 AGN.

During its 2005 outburst, \object{3C454.3} was the most luminous
object ever recorded in the optical band.  To explain such an
outburst, the jet power larger than $7 \times 10^{47}\,$erg s$^{-1}$
is required (see Table~\ref{tab:param}).  Is it feasible?  Noting that
the estimates of the black hole mass in this object give $\sim 4
\times 10^9 M_{\odot}$ \citep{Gu01}, we infer that the jet power is on
the order of the Eddington luminosity.  This, however, is at least by
a factor of few larger than the accretion luminosity, which in turn,
as determined from the optical luminosity of the thermal component
detected during the low state \citep{Smith88}, and after application
of the bolometric correction, is likely to be of the order
$10^{47}\,$erg s$^{-1}$.  \object{3C454.3} is in this respect not
exceptional among most powerful radio-loud quasars: powers of jets
larger than $10^{47}\,$erg s$^{-1}$ have been inferred for several
other quasars from analysis of the lobe energetics \citep{Rawlings91},
as well as from Chandra and HST observations of gamma-ray blazars
\citep{Tavecchio07}.

\acknowledgments This project was partially supported by Polish KBN
grant 5 P03D 00221 and NASA observing grant NNX07AB05G. This work was
also supported, in part, by the Department of Energy contract to SLAC
no.\ DE-AC3-76SF00515.  This research has made use of the NASA/IPAC
Extragalactic Database (NED) which is operated by the Jet Propulsion
Laboratory, California Institute of Technology, under contract with
the National Aeronautics and Space Administration.

\appendix

\section{Analytical approximations of the model parameters\label{app:approx}}
\subsection{Injection function}
Normalization factor $K_e$ of the electron injection function $Q$ can
be derived using approximate formulas for production of the X-ray
spectrum via the ERC process in the slow cooling regime
\citep[see][]{Moderski03}:
\be \nu_{x}L_{\nu_x} = {1 \over 2} [\gamma N_{\gamma}] |\dot
\gamma|_{ERC}(\theta_{obs}) m_e c^2 {\cal D}^4 \ee
where
\be |\dot \gamma|_{ERC}(\theta_{obs}) = {c\sigma_T \over m_ec^2}
u_{IR}' \gamma^2 \left({{\cal D} \over \Gamma_j}\right)^2 \ee
\be N_{\gamma} = Q {\Delta r \over c \Gamma_j} \ee
\be u_{ext}' = {4\over 3} \Gamma_j^2 u_{ext} \ee
and
\be {\cal D} = {1 \over \Gamma_j(1-\beta \cos{\theta_{obs}})} \ee
In the slow cooling regime the slope $p$ of the electron injection
function is $p=2 \alpha_x +1$ and for $\Delta r = r$ above equations
give
\be K_e = {3\over 2} {\nu_{x}L_{\nu_x} \over \sigma_T u_{IR} r}
{\Gamma_j \over {\cal D}^{4+2\alpha_x}} \left({\nu_{ext} \over
\nu_x}\right)^{1 - \alpha_x} \ee

In one of our models, the break in the injection functions is
introduced in order to get a better fit of the observed spectrum:
\be Q = K_e {1\over \gamma^p + \gamma_{br}^{p-q}\gamma^q }\ee
where $\gamma_{br}$ is the break energy and $q$ is the spectral index
of the injection function at high energy limit.

\subsection{Magnetic field intensity}
The ERC to synchrotron peak luminosity ratio
\be {L_{ERC} \over L_{syn}} = {u_{ext}' ({\cal D}/\Gamma_i)^2 \over
u_B'} \ee
gives us magnetic field intensity 
\be B' = {\cal D} \sqrt{{32 \over 3} u_{ext} {L_{ERC} \over L_{syn}}}
\ee
and magnetic energy flux
\be L_B = c u_B' \pi R^2 \Gamma_j^2 = \pi c u_B' r^2 (\theta_j
\Gamma_j)^2 \ee
where $u_B' = B^{\prime 2} /( 8 \pi)$ is magnetic energy density.
With known $B'$ we can estimate the maximum energy of injected electrons
\be \gamma_{max} \simeq 5.2 \times 10^{-4}
\sqrt{{\nu_{syn,max,obs}(1+z) \over B' {\cal D}}} \ee

\subsection{Electron energy density}
Due to light travel effects, sources moving with relativistic speeds
are seen on the sky as stretched by a factor ${\cal D} \Gamma_j$,
which means that only a fraction $1/({\cal D} \Gamma_j)$ of particles
is seen at a given instance to be enclosed within the distance range
$\Delta r$.  Hence the volume of the jet segment into which electrons
are injected at the 'observed' rate $Q$ is $\pi R^2 \lambda$, where
$\lambda = \Delta r /({\cal D}\Gamma_j)$.  Amount of energy injected
into the segment during its propagation through the $\Delta r$ zone is
\be E_{e,inj}' = {\Delta r \over c \Gamma_j} \int{Q \gamma m_ec^2 \,
d\gamma} \ee
and energy density of injected electrons is
\be u_{e,inj}'(r_0 + \Delta r= 2 r_0) = { E_{e,inj}' \over \pi R^2
\lambda'} = {{\cal D} \over \Gamma_j} {\int{Q \gamma m_ec^2 \,
d\gamma} \over \pi c R^2} = {m_e c {\cal D} \Gamma_j \int{Q \gamma \,
d\gamma} \over 4\pi r_0^2 (\theta_{obs}\Gamma_j)^2} \ee
where $\lambda'=\lambda \Gamma_j$.

\subsection{Energy dissipation efficiency}
In the proton inertia dominated jets acceleration of electrons is
powered by protons and we have
\be u_{e,inj}' = \eta_e u_p' (\bar \gamma_p -1) \ee
where $(\bar \gamma_p -1) \ll 1$ is the fraction of proton bulk
kinetic energy converted to the 'thermal' proton energy called
hereafter the efficiency of energy dissipation, and $\eta_e$ is the
fraction of proton 'thermal' energy tapped by electrons. Condition of
having matter dominated jet implies $u_p' > u_B'$, and combining this
with previous equation gives
\be (\bar \gamma_p -1) < {u_{e,inj}' \over u_B' \eta_e}
\label{eq:appgp} \ee

\subsection{Pair content}
Using definition of particle energy densities ($u=nmc^2\bar\gamma$)
and noting that $\bar \gamma \gg 1$ (throughout our paper, $\gamma
\equiv \gamma_e$) and $\bar \gamma_p -1 \ll 1$ we obtain the pair
content
\be {n_e' \over n_p'} = {m_p \over m_e} {\bar \gamma_p -1 \over \bar
\gamma} < {m_p \over \eta_e \bar \gamma m_e} {u_{e,inj}' \over u_B'}
\ee
where inequality (\ref{eq:appgp}) was used and $\bar \gamma \equiv
\int{Q\gamma \, d\gamma}/\int{Q \, d\gamma}$.

\subsection{Toroidal vs. turbulent magnetic field}
We assumed in the paper that magnetic field is dominated by the
toroidal component.  This assumption can be verified as follows.  For
$u_{B,tor}' >> u_{B,turb}'$, $u_{B,tor}' \simeq u_{B,tot}' \equiv
u_B'$ and
\be u_{B,tor}' \simeq {u_B' \over u_{e,inj}'} u_{e,inj}' = \eta_e
{u_B' \over u_{e,inj}'} u_p'(\bar \gamma_p -1) \ee
For $u_{B,turb}' \simeq \eta_B u_p'(\bar \gamma_p - 1)$ this gives
\be {u_{B,tor}' \over u_{B,turb}'} = {\eta_e \over \eta_B} {u_B' \over
 u_{e,inj}'} \ee
\bigskip

Note that all formulas which involve a Doppler factor apply for
'mono-Doppler' sources only.  In the case of conically diverging
jets, the observed radiation is contributed by jet portions moving
relative to the line of sight at different angles and then analytical
estimations differ significantly from numerical results.  This in
particular concerns the quantity $K_{\rm e}$ because of its strong
dependence on ${\cal D}$.  However, for $\theta_{obs} \sim \theta_j
\sim 1/\Gamma_j$, still reasonable analytical estimates are achievable
if using ${\cal D} = 1.5 \Gamma_j$, instead of ${\cal D}= \Gamma_j$.

\section{The millimeter photosphere\label{app:photo}}
Optically thin synchrotron spectrum in \object{3C454.3} and other
quasar hosted blazars is produced by electrons in the fast cooling
regime.  In this regime an electron distribution is steepened due to
radiative losses, and for a single-power-law injection function, $Q
\sim \gamma^{-p}$, the electrons reach a distribution with the index
$s= p+1$.  Below we provide estimation of the millimeter photosphere
distance, assuming $p=2$.  For such a source the synchrotron-self
absorption opacity $\tau(\nu_{abs}')$ is at $\nu_{\rm a}'$ equal to 1
for
\be R_{mm} = 2.7 \times 10^{-15} {\nu_{\rm a}^{\prime 7/2} \over c_n
B^{\prime 5/2}} \, {\rm [cm]}\ee
where $n_{\gamma} = c_n \gamma^{-3}$ is the electron density energy
distribution.

Noting that 
\be c_n = {C_{N} \over V'} = {C_N \Gamma_j^2 {\cal D} \over \pi r^3
(\Delta r/r) (\Gamma_j \theta_j)^2} \ee
where $C_N: N_{\gamma} = C_N \gamma^{-3}$, and that
\be N_{\gamma} = {\int_{\gamma} Q \, d\gamma \over |\dot\gamma|_{tot}}
\ee
where for $L_{ERC} > L_{syn}$
\be |\dot\gamma|_{tot} \simeq {16 c \sigma_T \gamma^2 \Gamma_j^2
u_{ext} \over 9 m_e c^2} \ee
we obtain, for $\theta_{obs} \Gamma_j=1$ and $\Delta r=r$,
\be R_{mm} \simeq 1.9 \times 10^7 {{\cal D}^{9/5} \over
\Gamma_j^{7/5}} {B_0' r_0 \over (u_{ext,in}r_{in}^2)^{2/5}} {K_e^{2/5}
\over [\nu_{\rm a,obs}(1+z)]^{7/5}} \, {\rm [cm]} \ee
and $r_{mm} = R_{mm}/\Gamma_j$.  For $\nu_{\rm a,obs} = 3 \times
10^{11}$ Hz ($\lambda_{\rm a,obs} = 1$mm) and parameters of the Model
1 (see Table~\ref{tab:param}), this gives $R_{mm} \simeq 1.4 \times
10^{18}\,$cm and $r_{mm} \simeq 2.8 \times 10^{19}\,$cm.

\bibliography{ms}

\clearpage

\clearpage

\clearpage

\newpage

\begin{table}
  \caption{The model parameters.}\label{tab:param}
  \begin{center}
    \begin{tabular}{lcccc}  
    \tableline\tableline
    Parameter & Model 1 & Model 2 & Model 3 & Model 4 \\
    \tableline\tableline
    $\gamma_{\rm min}$ & $1$ & $1$ & $1$ & $1$ \\
    $\gamma_{\rm br}$ & --- & --- & --- & $80$ \\
    $\gamma_{\rm max}$ & $4 \times 10^3$ & $4 \times 10^3$ & $4 \times 10^3$ & $9 \times 10^3$ \\
    $p$ & $2.0$ & $2.0$ & $2.0$ & $1.7$ \\
    $q$ & $2.0$ & $2.0$ & $2.0$ & $2.5$ \\
    $K_{\rm e}\, [\rm s^{-1}]$ & $3.0 \times 10^{49}$ & $1.5 \times 10^{49}$ & $2.3 \times 10^{49}$ & $3.0\times 10^{48}$ \\
    $\Gamma_{\rm j}$ & $20$ & $20$ & $20$ & $20$ \\
    $\theta_{\rm j}$ \, [rad] & $0.05$ & $0.05$ & $0.05$ & $0.05$ \\
    $\theta_{\rm obs} \, [{\rm rad}]$ & $0.05$ & $0.05$ & $0.05$ & $0.05$ \\
    $r_0= \Delta r_0 \, [{\rm cm}]$ & $10^{19}$ & $2 \times 10^{19}$ & $2 \times 10^{19}$ & $2 \times 10^{19}$ \\
    $B_0 \, [{\rm G}]$  & $1.4$ & $0.50$ & $0.63$ & $0.27$ \\
    $r_{\rm in} \, [{\rm cm}]$ & $10^{19}$ & $10^{19}$ & $10^{19}$ & $10^{19}$ \\
    $u_{\rm IR}(r_{\rm in}) \, [{\rm erg \, cm^{-3} \, s^{-1}}]$  & $1.24 \times 10^{-4}$ & $1.24 \times 10^{-4}$ & $1.24 \times 10^{-4}$ & $1.24 \times 10^{-4}$ \\
    $h \nu_{\rm IR} \, [{\rm eV}]$ & $0.34$ & $0.34$ & $0.34$ & $0.34$ \\
    \tableline
    $u'_{\rm e,inj}(2r_0) \, [{\rm erg \, cm^{-3} \, s^{-1}}]$ & $3.25 \times 10^{-3}$ & $4.06 \times 10^{-4}$ & $6.22 \times 10^{-4}$ & $1.95 \times 10^{-4}$ \\
    $u_B'(2r_0) \, [{\rm erg \, cm^{-3} \, s^{-1}}]$         & $1.95 \times 10^{-2}$ & $2.49 \times 10^{-3}$ & $3.95 \times 10^{-3}$ & $7.25 \times 10^{-4}$ \\
    $L_{\rm j} > L_B \, [{\rm erg \, s^{-1}}]$  & $7.35 \times 10^{47}$ & $3.75 \times 10^{47}$ & $5.96 \times 10^{47}$ & $1.10 \times 10^{47}$ \\
    $\bar \gamma_p - 1 <$ & $0.17/\eta_e$ & $0.16/\eta_e$ & $0.16/\eta_e$ & $0.26/\eta_e$ \\
    $\bar \gamma$ & $8.3$ & $8.3$ & $8.3$ & $13.0$ \\
    $n_e/n_p <$ & $37.5/\eta_e$ & $35.3/\eta_e$ & $35.3/\eta_e$ & $37.9/\eta_e$ \\
    $u_{B_{tor}}' / u_{B_{turb}}'$ & $5.88 \, \eta_e/\eta_B$ & $6.25 \, \eta_e/\eta_B$ & $6.25 \, \eta_e/\eta_B$ & $3.72 \, \eta_e/\eta_B$ \\
    \tableline\tableline
   \end{tabular}
  \end{center}
\end{table}

\end{document}